\documentclass[12pt,epsfig]{article}
\usepackage{graphicx}
\usepackage{amssymb}
\newcommand{\beq}[1] {\begin{equation}\label{#1} }
\newcommand{\eeq} {\end{equation} }

\newcommand{\bea}[1]{\begin{eqnarray}\label{#1} }
\newcommand{\eea}{\end{eqnarray}}

\newcommand{\vn}{{\vec{n}}}

\newcommand{\si}{\sigma}

\begin{document}

\vspace*{-0.5cm}
\begin{flushright}
OSU-HEP-04-08\\
SU-4252-799 \\
Fermilab-Pub-04/127
\end{flushright}
\vspace{0.5cm}

\begin{center}
{\Large
{\bf Single Kaluza Klein Production \\ in Universal Extra Dimensions} }

\vspace*{1.5cm}
 C. Macesanu\footnote{Email address:  mcos@pas.rochester.edu}$^{,\dag,\ddag}$,
S. Nandi\footnote{Email address: shaown@okstate.edu}$^{,\dag,\sharp}$
and M. Rujoiu\footnote{Email address: mrujoiu@okstate.edu}$^{,\dag,*}$

\vspace*{0.5cm}
$^{\dag}${\it Department of Physics, Oklahoma State University\\
Stillwater, Oklahoma, 74078\\}

$^{\ddag}${\it Department of Physics, Syracuse University\\
Syracuse, New York 13244}

$^{\sharp}${\it Fermi National Accelerator Laboratory\\
P.O. Box 500, Batavia, Il 60510}\footnote{ S.N.
was participant in the Summer Visitor Program at Fermilab.}

$^{*}${\it Institute of Space Sciences, Bucharest-Magurele\\
Romania, 76900\\}
\end{center}

\begin{abstract}
In the universal extra dimensions models, Kaluza Klein excitations
of matter are generaly produced in pairs. However, if matter
lives on a fat brane embedded in a  larger space, gravity-matter interactions
do not obey KK number conservation, thus making possible the production
of single KK excitations at colliders. We evaluate the production
rates for such excitations at the Tevatron and LHC colliders, and look
for ways to detect them.

\end{abstract}

\section{Introduction}

In ADD-type models \cite{ADD},
the 4D universe we live in is viewed as a brane in a space
with 4+N dimensions.  Gravity can propagate in all dimensions, while matter
fields (and the gauge bosons) are restricted on the 4D brane.  As a consequence,
the gravity field will appear to an observer on the Standard Model brane
as consisting of a massless graviton plus an infinite number of
Kaluza-Klein modes, with mass spacing proportional to the inverse length of
the extra dimensions. Based on arguments of naturalness (that is, the
fundamental gravity scale in 4+N dimensions should be close to the electroweak
scale), the mass splitting of the KK gravitons range from $10^{-3}$ eV order (for N = 2)
to MeV order (for N = 6).    

One might consider modifying the above scenario by letting the matter
fields propagate into one or more extra dimensions. Then ordinary matter
will also have KK exciations with masses of order $1/R$. Noting that no such
excitations have been seen at colliders, one should conclude that $R$ should be
at least of inverse TeV order. However, we then have to change our picture about
gravity, by either assuming  that the extra dimensions are asymmetrical
\cite{asym_ed}
(some of sub-millimeter order, in which only gravity propagate, and some of order inverse TeV,
in which matter can also propagate) or, if we keep all 
compact dimensions of TeV$^{-1}$ size, by 
reintroducing some hierarchy between the 
gravity and electroweak scales.

Perhaps a more satisfying picture would be one in which the gravity sector
is unaffected (the same as in ADD), but matter propagates only in part of
the extra dimensions. In other words, the 4D brane on which
Standard Model matter lives is endowed 
with a finite thickness in the extra-dimensions, resulting in what 
is known as the `fat brane' scenario \cite{rigolin}. We also assume that
all the Standard Model matter fields live on this fat brane, like in 
Universal Extra Dimensions (UED) models \cite{acd}.  

The consequences of this scenario for phenomenology are quite interesting.
In the standard UED models, the first level KK excitations are stable
(due to KK number conservation and mass degeneracy at tree level),
therefore they are rather hard to see
at colliders. In the fat brane scenario, KK number conservation is
broken in the gravity-matter interactions, so these first level KK 
excitations can decay by radiating a graviton. Since there is a large
number of gravitons with mass less than the TeV scale, the gravitational 
decay width of these particles is large enough that the decay takes place
inside the detector, and the experimental signal will be jets plus missing
energy \cite{mmn}. Moreover, if radiative corrections alter the massess
of the first level excitations \cite{ggh,cms1}, 
an even richer phenomenology can ensue, with
photons as well as leptons as end products of the decays of the 
heavy KK excitations \cite{cms2,mmn2}.

However, besides consequences for the decay of KK matter excitations, gravity
interactions can also mediate production of a single KK excitation in 
colliders. This is unlike the usual case in the UED scenario, where
KK matter excitations are produced in pairs. Since production of a pair
of heavy particles is often kinematically restricted, one might 
then envision the case where it will
be easier to produce a single KK excitation (provided that the gravitational
interaction is strong enough). The study of this possibility is the 
purpose of this paper.

The outline of the paper is as follows. In the next section we will give a 
brief overview of the model under consideration, together with the
analytic expressions for the square amplitudes of the relevant processes.
In section 3 we present the cross-sections for the production of one
jet + one excited KK state in the UED model with a fat brane, at
the Tevatron and Large Hadron Collider (LHC). We also
look at the dependence of the signal cross-section (jets + missing energy) on
cuts on relevant observables, and compare the signal with the Standard Model
background. We end with conclusions.

\section{Model description}

In our scenario, matter propagates in one compact extra dimension, 
which, in order to accomodate the chiral fermions of the Standard
Model (SM),
is an $S_1/Z_2$ orbifold with a radius $R$ of order inverse TeV. As a 
consequence of orbifolding, the SM scalar and gauge boson fields aquire
each one KK excitation tower (the modes being even under the $Z_2$ parity),
while each SM fermion field aquire two KK excitation towers (the tower
correponding to the left handed fermions being different than the tower
associated with the right handed fermions) \cite{acd,mmn}. The masses of the 
particles in these towers are, at tree level, multiples of $M = 1/R$.

The space in which matter propagates is embeded as a fat brane
(\cite{rigolin}) in a larger space of
$4 + N$ dimensions where gravity propagates. The length $r$ of these extra 
dimensions is given in terms of the fundamental Plank scale $M_D$ by the
ADD relation:
$$ M_{Pl}^2 = M_D^N \left({r \over 2 \pi}\right)^{N+2} ,$$
where $M_{Pl}$ is the usual Plank scale in 4 dimensions.
The interactions between matter and gravity in this model have been worked out
in \cite{grv-pap}. The Feynman rules for matter-gravitation interactions are
quite similar to those obtained for the case of matter propagating into
4 dimensions only \cite{hlz}, except that each vertex is multiplied
by a form-factor
$${\cal F}^c_{lm|n} \sim { 1\over \pi R}
\int_0^{\pi R} dy \cos \left({l y \over R}\right)
 \cos \left({m y \over R}\right) \exp \left(2 \pi i {n y \over r}\right)
$$
which describe the superposition in the fifth dimension of the graviton
and matter  wave functions. 

Having matter and gravity propagate on different scales in the fifth 
dimension means that KK number
conservation will not hold for gravity-matter interactions. This has two
consequences: first is that the first level KK excitations of matter will
decay to SM matter radiating a graviton. Since there is a whole KK graviton
tower which can mediate this decay, the decay width (evaluated in 
\cite{grv-pap}, for example ) is large enough that any KK excitation of matter
produced at colliders will decay inside the detector\footnote{In this paper
we will consider only the gravitational decays of the matter KK excitations,
neglecting decays due to mass splittings.
This is justified by the fact that the gravitational decay width increases
as (mass)$^3$, and the KK states used in our analysis
are rather heavy.}. 
The phenomenology
resulting from these decays has been studied in \cite{mmn}.

The second consequence is  that it is also possible to produce 
a single KK excitation of matter (unlike the usual UED case, where
these excitations are produced in pairs). This will take place through the
exchange of gravitons; again, since there is a large number of graviton
excitations, the contribution coming from all of these can be big enough
to offset the $1/M_{pl}$ scale coupling of the individual graviton. The
phenomenology resulting from this scenario leads to interesting signals for the 
extra dimensions, and is discussed in the following.

\subsection{Single KK production}
We give here the analytic expressions for the squared amplitudes for the 
processes contributing to the production of one KK excited state of matter.
We use the following modified Mandelstam variables:
$$ s = 2 p_1 p_2 \ ,  \ t = -2 p_1 p_3 \ , \ u = -2 p_1 p_4 $$
where $p_1, p_2$ are the momenta of the initial state partons, and
$p_3, p_4$ the momenta of the final state ones. $D(s), D(t)$ and $D(u)$ are
the summed graviton propagators (see next section) in the $s,t$ and $u$ 
channels.

With these notations, the amplitudes for the relevant proceses are
(we denote by a $^*$ the KK excitation of a SM particle):

\begin{enumerate}

\item for $q \bar{q} \rightarrow g^{*} g$ ($s$ channel only):
\begin{eqnarray*}
\bar{\sum}|M(q \bar{q}\rightarrow g^{*}g)|^{2}=
\frac{1}{3}tu(t^{2}+u^{2})*D(s)^{2}
\end{eqnarray*}

\item for $g g \rightarrow g^{*}g $ ($s,t$ and $u$ channels):
\begin{eqnarray*}
\bar{\sum}|M(g g \rightarrow g^{*}g)|^{2}=
\frac{1}{18}[(t^{4}+u^{4})*D(s)^{2}+(s^{4}+u^{4})*D(t)^{2}+\\
(s^{4}+t^{4})*D(u)^{2}+2u^{4}*D(s)*D(t)+
2t^{4}*D(s)*D(u)+2s^{4}*D(u)*D(t)]
\end{eqnarray*}

\item for $q g \rightarrow q g^{*} $ and
$\bar{q} g \rightarrow \bar{q} g^{*} $ ($t$ channel only):
\begin{eqnarray*}
\bar{\sum}|M(q g \rightarrow q g^{*})|^{2}=
-\frac{1}{12}s u (s^2+u^2)*D(t)^{2}
\end{eqnarray*}

\item for $g g \rightarrow q^{*}\bar{q} $ ($s$ channel):
\begin{eqnarray*}
\bar{\sum}|M(g g \rightarrow q^{*} \bar{q})|^{2}=\frac{1}{96}
tu(t^{2}+u^{2})*D(s)^{2}
\end{eqnarray*}

\item for $q \bar{q} \rightarrow q^{*} \bar{q}$ and 
$q \bar{q} \rightarrow q \bar{q}^*$:
\begin{eqnarray*}
  \bar{\sum}|M(q \bar{q}\rightarrow q^{*} \bar{q})|^{2}=\frac{1}{256}
  [(s^{4}-10 s^{2} t u +32t^2 u^{2})*D(s)^{2}\\
  +(t^{4}-10 t^{2} s u +32 s^2 u^{2})*D(t)^{2}
  -2 u^{2} (4 u^{2}+ 9 s t)*D(s)*D(t)]
\end{eqnarray*}

\item for $q q \rightarrow q^{*}q $ and 
 $\bar{q} \bar{q} \rightarrow \bar{q}^{*}\bar{q} $:
\begin{eqnarray*}
\bar{\sum}|M(q q \rightarrow q^{*} q)|^{2}=
\frac{1}{256}[(t^{4}-10 t^{2} s u +32 s^2 u^{2} )*D(t)^{2}+\\
(u^{4}-10 u^{2} s t +32 s^2 t^2)*D(u)^2-
2 s^{2} (4 s^{2}+ 9 t u)*D(t)*D(u)]
\end{eqnarray*}

\item for $q g \rightarrow q^{*}g $:
\begin{eqnarray*}
\bar{\sum}|M(q g \rightarrow q^{*} g)|^{2}=
-\frac{1}{24}s u (s^{2}+u^{2})*D(t)^{2}
\end{eqnarray*}

\end{enumerate}

Also, the averaged squared amplitudes for scattering processes with
different flavor quarks in the initial state 
($q \bar{q}' \rightarrow q^{*} \bar{q}'$ and
$q q' \rightarrow q^{*}q' $) are given by the $t$ channel contributions only
from the corresponding expressions (5) and (6) above.

\subsection{Summed propagator}

The effective propagator for gravity mediated KK production is obtained
by summing over all graviton excitations up to a cut-off $M_D$. Thus,
\beq{grv_sum}
 D(s) = \kappa^2 \ \sum_\vn {\cal F}_{0|n_5} \ {i \over s - m_\vn^2 } \
({\cal F}^c_{1|n_5})^* \ , \eeq
where $\kappa = \sqrt{16 \pi / M_{Pl}^2} $ is  the 4D gravitational 
coupling constant, 
and ${\cal F}_{0|n_5}$ and ${\cal F}^c_{1|n_5}$ are form factors
describing the interaction of the gravitons with the matter excitations
on the brane (\cite{grv-pap}). Note that terms in numerator of the propagator
for a single graviton proportional to $p^{\mu}, p^{\nu}$ 
(\cite{hlz}) drop out due to the fact that one end of the propagator
couples always to two massless fermions. Hence the Lorenz structure of
the propagator for the spin-2 massive graviton is simply:
$$
B(k)_{\mu \nu,\rho \si} = \left(
\eta_{\mu \rho}\eta_{\nu \si} +
\eta_{\mu \si}\eta_{\nu \rho} -
{2 \over 3} \eta_{\mu \nu}\eta_{\rho \si} \right) D(k^2) \ .
$$

\begin{figure}[h!] 
\centerline{
   \includegraphics[height=3.in]{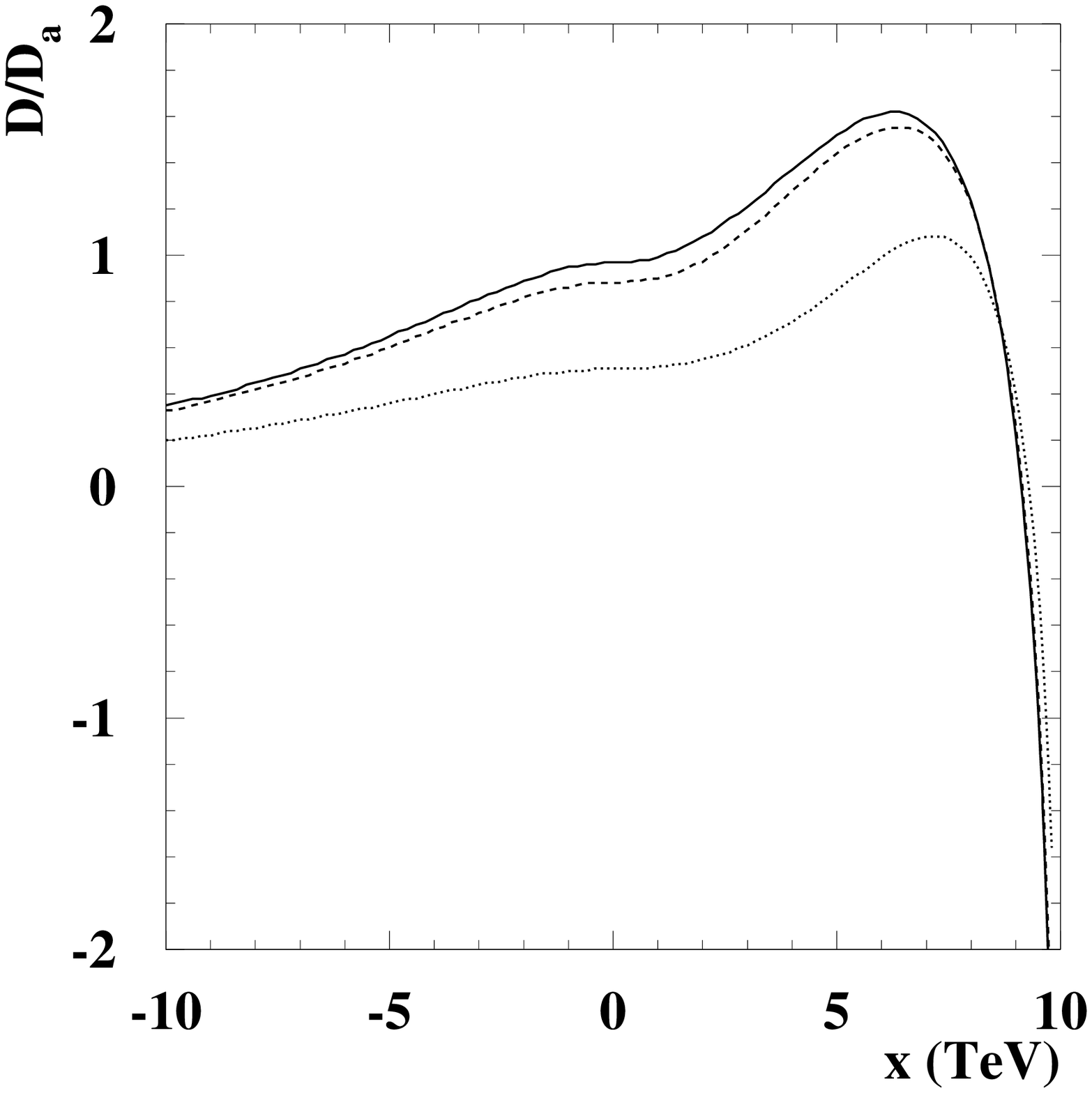}
   \includegraphics[height=3.in]{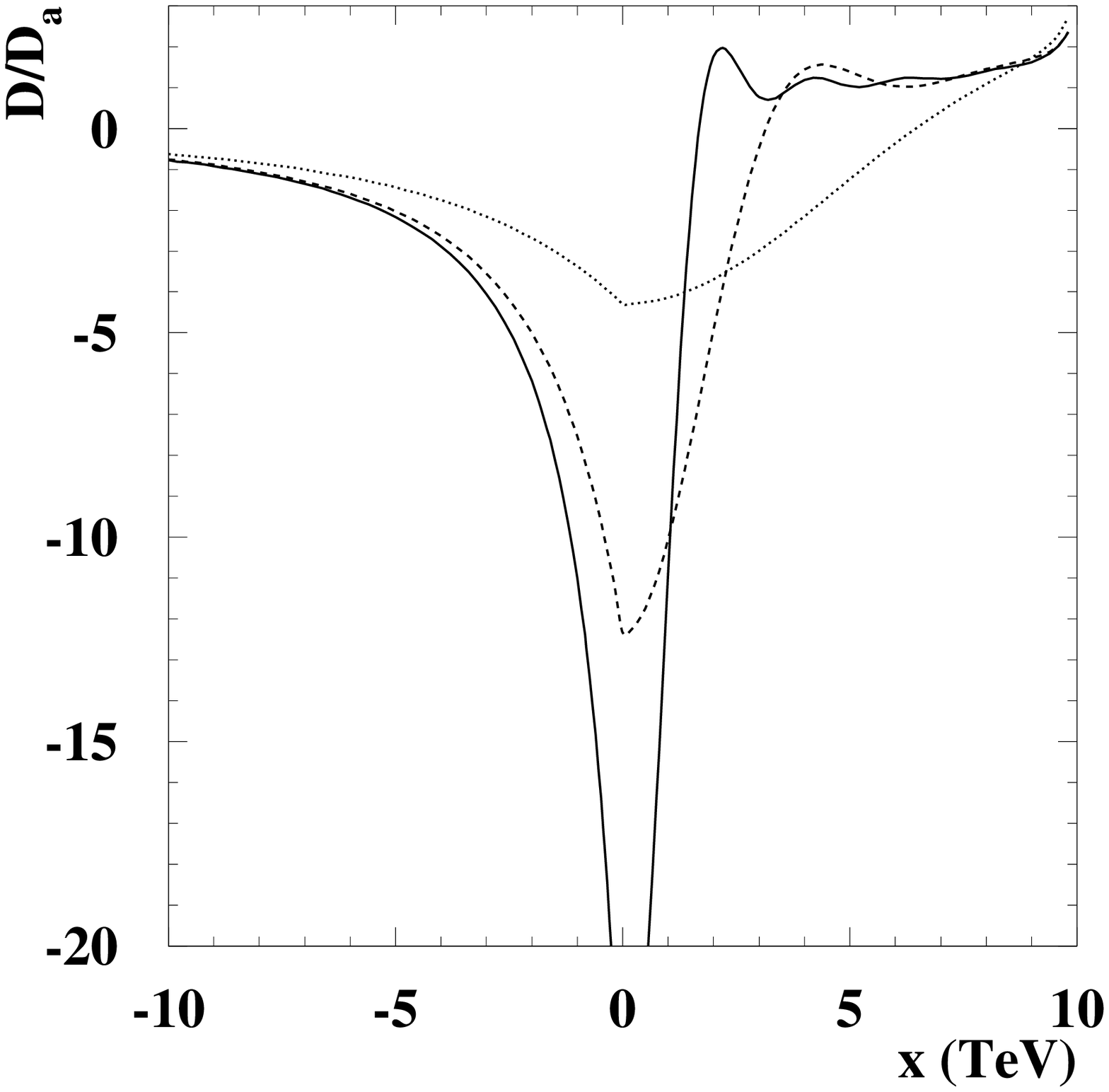}
   }
\caption{ Ratio of values for exact propagator versus analytic approximation,
for $N = 6$ (left) and $N=2$ (right) extra dimensions. 
Here $M_D = 10$ TeV. Lines correspond to values for $M$ of 1 TeV (straight)
2 TeV (dashed) and 5 TeV (dotted line). On the horizontal axis is 
$x = \hbox{sign(s)} \sqrt{|s|}$. }
\label{gravp}
\end{figure}

In the limit where $s, M \ll M_D$, the sum (\ref{grv_sum}) has
been evaluated in \cite{grv-pap}:
\beq{grv_app}
D(s) \ \simeq \ 
V_{N-1} \
{ 16 \over N-3}  {M \over M_D^5}
{2 \sqrt{2} \over \pi^2} \
\int_0^{\pi M_s/M}  {\sin x \over 1-x^2/\pi^2} dx  \ .
\eeq
For the purposes of this paper, we evaluate (\ref{grv_sum}) numerically, for
each value of $s$ and $M_D$. It turns out that the approximate result
(\ref{grv_app}) it is reasonably accurate, 
except for the cases $N = 2,3$ (when strictly speaking it is not applicable). 
In Fig. \ref{gravp} we print the ratio
between the exact summed propagator (obtained by numerical integration) and
the analytic approximation (\ref{grv_app}); negative values of $s$ are
applicable for the case of $t$ and $u$ channel scattering.

\section{Results}
\begin{figure}[t!] 
\centerline{
   \includegraphics[height=3.in]{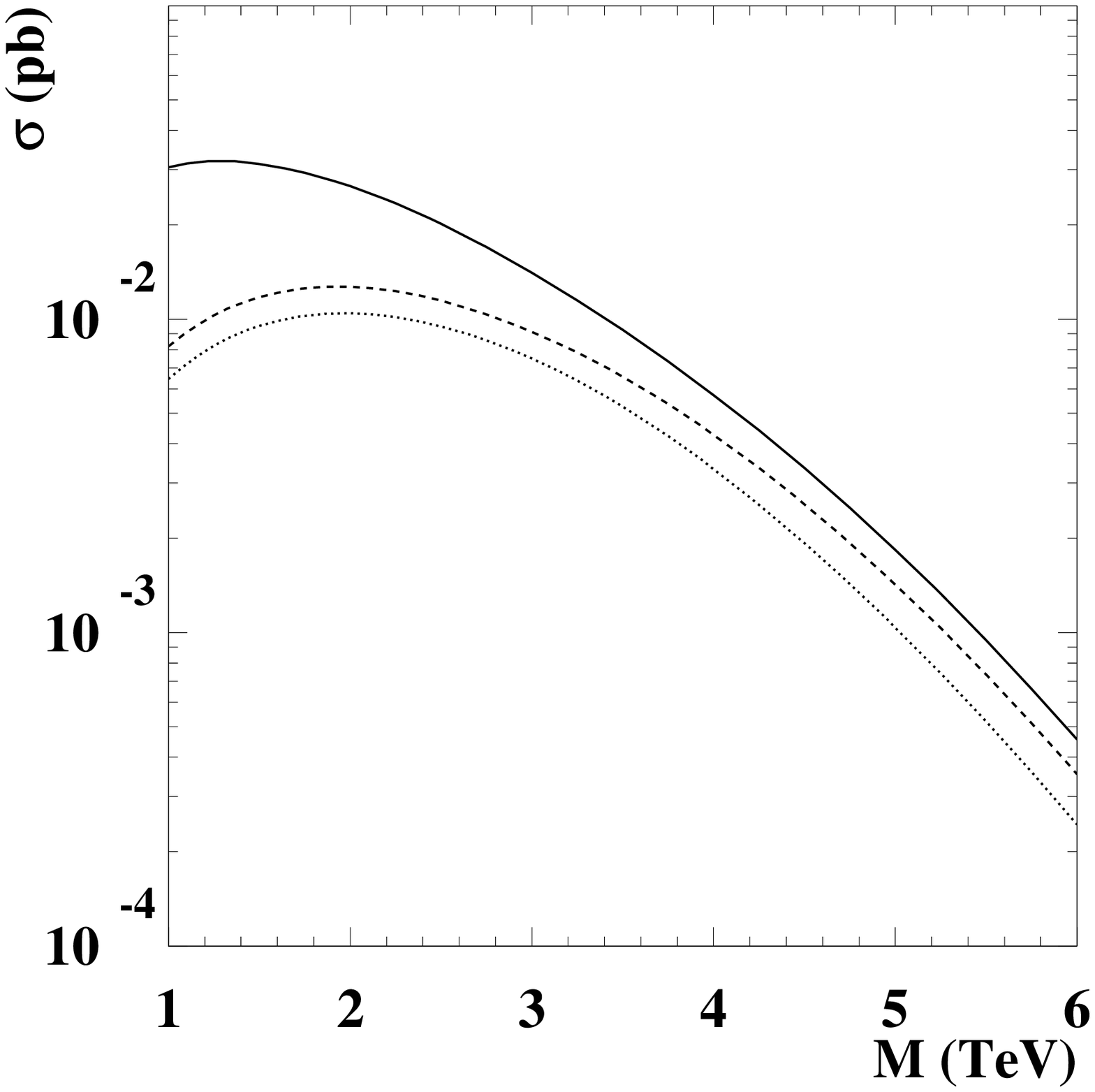}
   \includegraphics[height=3.in]{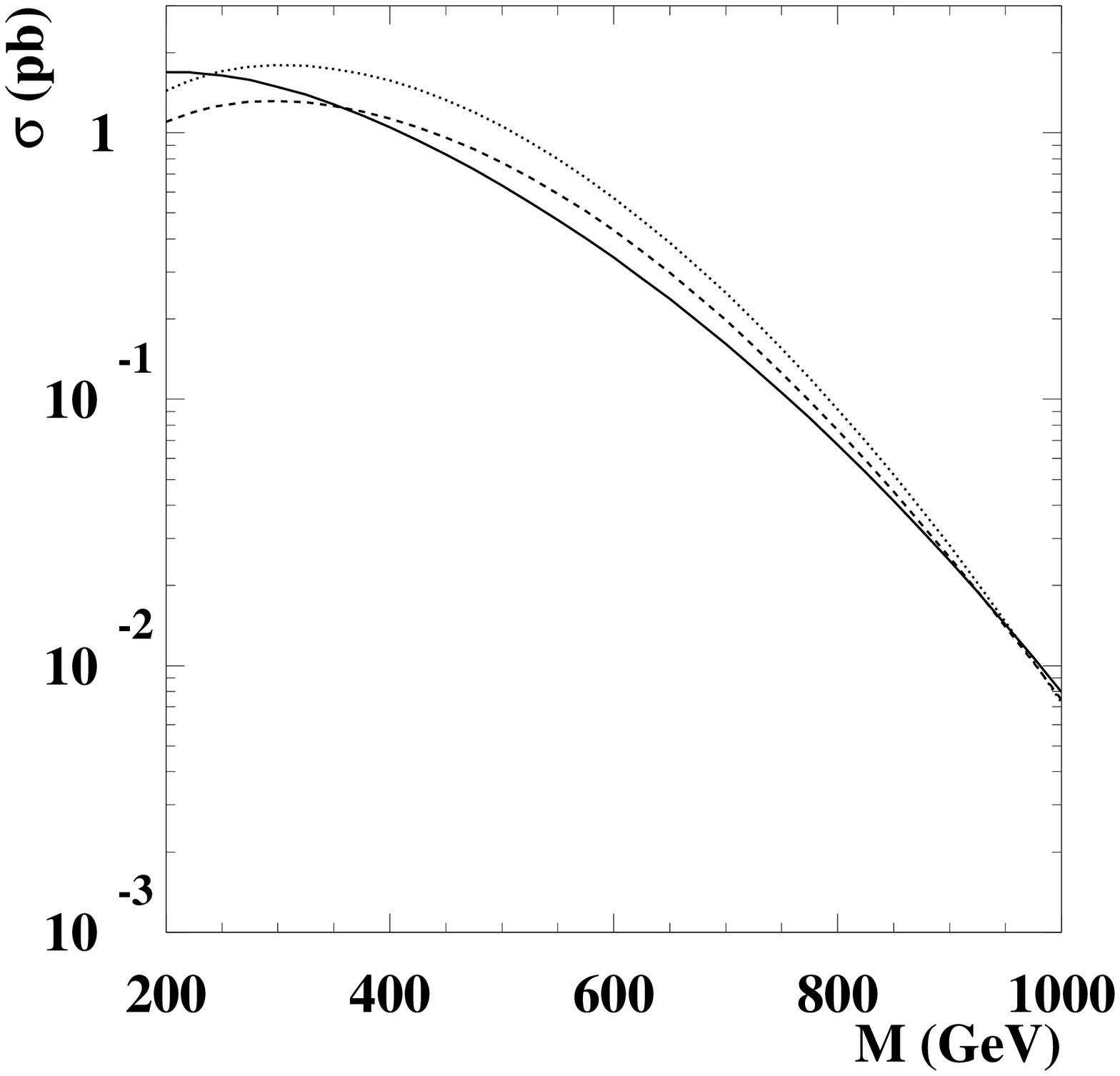}
   }
\caption{Production cross-sections at the LHC (right) and Tevatron RunII (left),
for $N = 2$ (straight line), 4 (dashed line)
and 6 (dotted line) extra dimensions.}
\label{prod_crs}
\end{figure}

The cross-sections for the gravity mediated production of one KK excitation
at LHC and Tevatron Run II are given in Fig. \ref{prod_crs}. As one can see,
 due to the fact that only a 
single heavy particle has to be produced, the reach of these machines has
the potential to be quite large.
Note, however, that the magnitude of the cross-section depends 
critically on the magnitude
of the fundamental Planck scale $M_D$ (the value of this parameter 
in Fig. \ref{prod_crs} is
set to $M_D = 10$ TeV for the LHC simulation and  $M_D = 1.5$ TeV for the
Tevatron one). 
This can be understood from Eq. (\ref{grv_app}); since the propagator depends
on the 5th power of $M_D$, this means that the production cross-section will
scale as $(1/M_D)^{10}$. Therefore, a doubling of the fundamental gravity scale
will reduce this cross section by roughly three orders of magnitude.

From Fig. \ref{prod_crs}, 
we see then that, at the LHC, there is a significant signal if the $M_D$ 
scale is around 10 TeV or lower, for masses of the KK excitations of matter
as high as 5 TeV. However, in order to make a discovery, one has to see the
signal over the Standard Model background. Since in this case the final state
consist of two partons and the graviton produced in the decay of a heavy KK
state,  the experimental signal will be two jets plus missing energy.
The SM background will then get contributions from Z + 2 jets processes (where
Z decays to $\nu \bar{\nu}$ or $\tau \bar{\tau}$ pairs), W + 2 jets (with the
lepton from the W decay unidentifiable), $t \bar{t}$ production, with
one top decaying to $b \bar{\nu} l$ and unidentified lepton and jets, 
and QCD multijet production with mismeasured
$\not{E_T}$. In order to be able to eliminate this background, one must use
cuts on the physical observables.

\begin{figure}[b!] 
\centerline{
   \includegraphics[height=3.in]{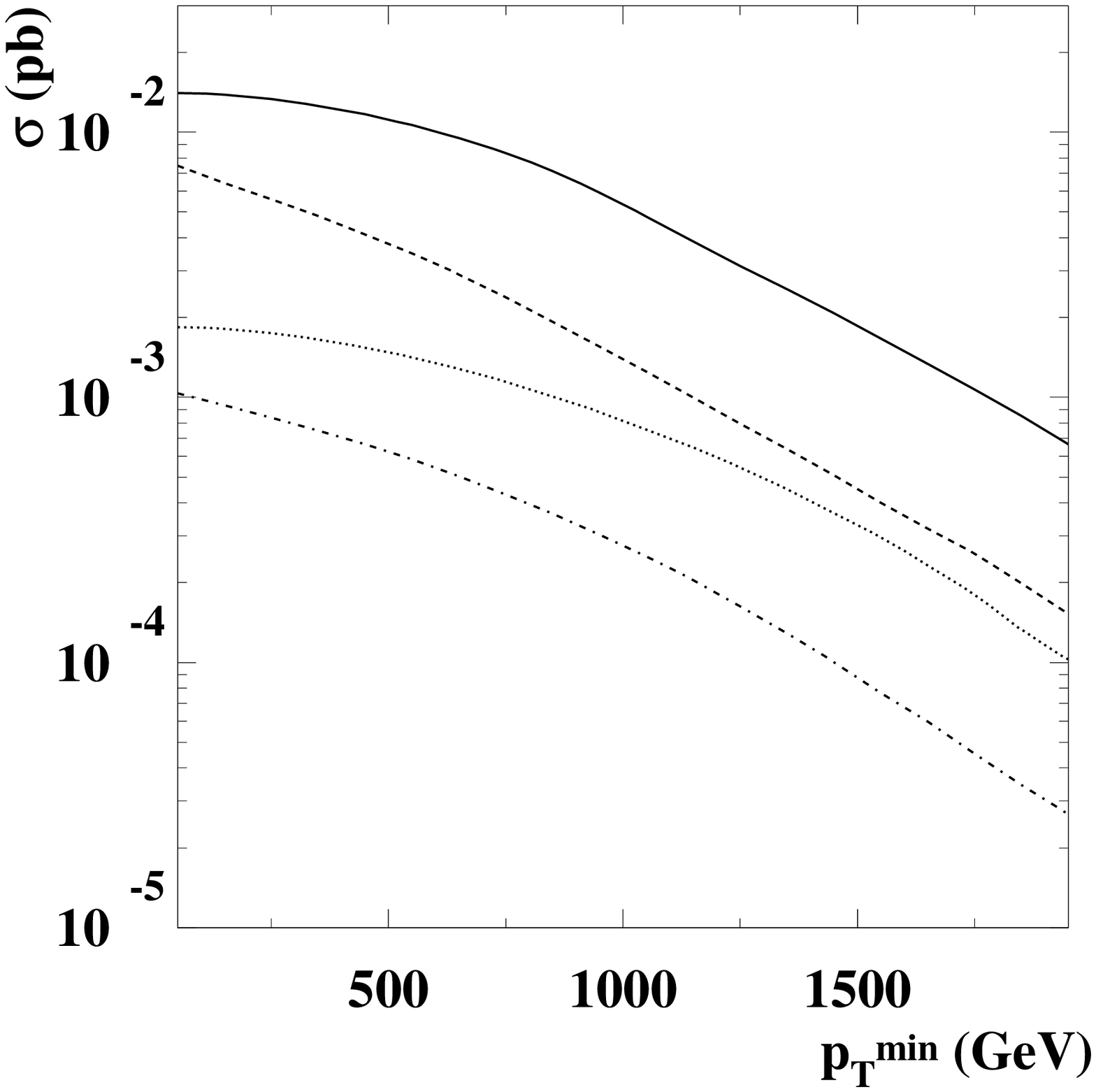}
   \includegraphics[height=3.in]{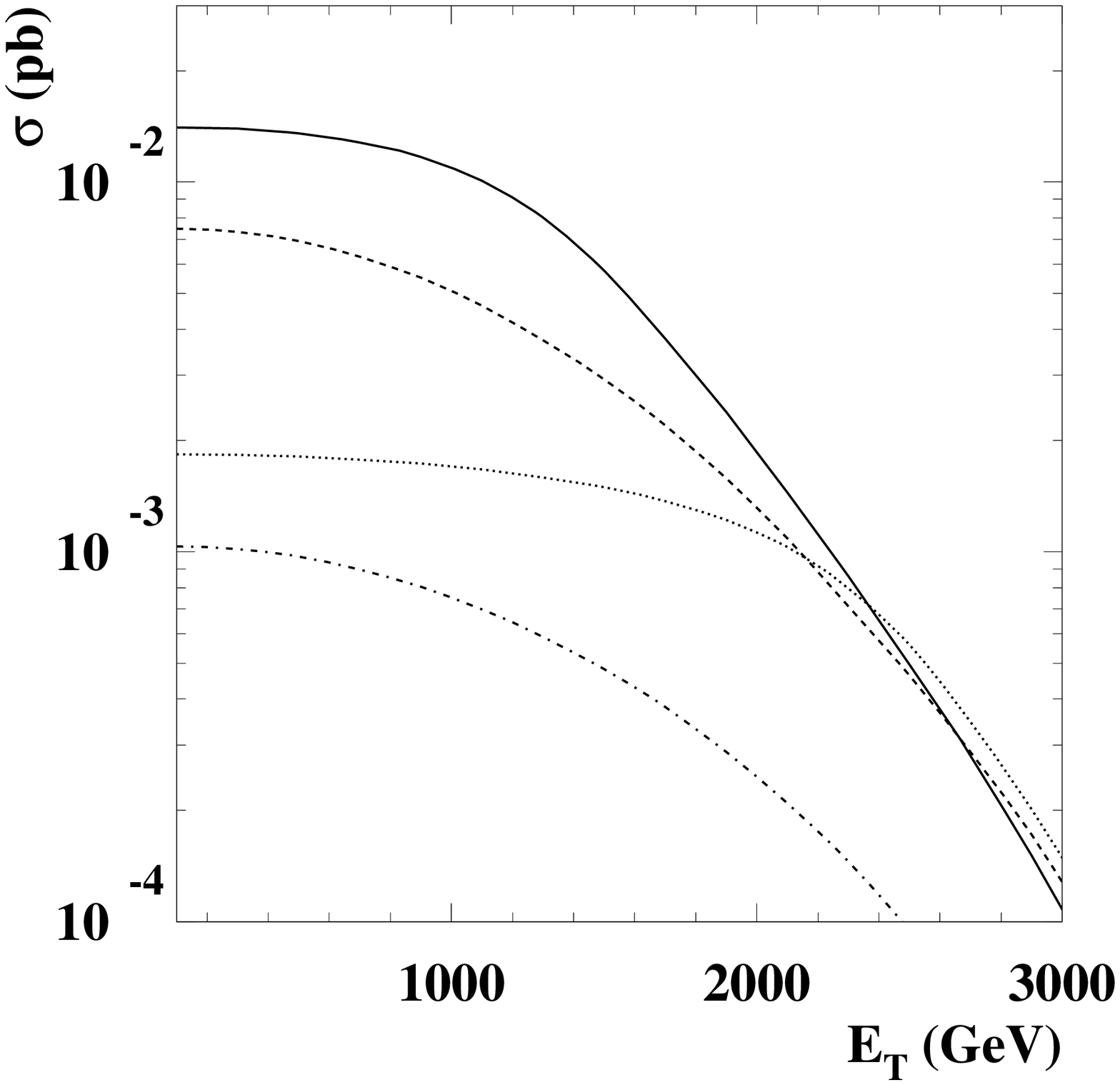}
   }
\caption{Cross section as function of cuts on the minimum $p_T$ of jets (left)
and missing energy $\not{E_T}$ in an event (right). Straight and dashed line
correspond to $M = 3$ TeV ($N = 2$ and $N=6$ respectively) while
dotted and dash-dotted lines correspond to $M = 5$ TeV 
(again, for $N = 2$ and $N=6$). }
\label{ptcuts}
\end{figure}

It is therefore interesting to consider our signal's dependence on cuts
on jet transverse momentum and the missing energy in an event. These 
distributions are presented in Fig. \ref{ptcuts}. We see that due to the
large mass of the KK particle being produced, the transverse momentum of the
jets is quite large. This also holds even more for the missing energy; it is
interesting to note that due to the fact that the invisible momentum goes all
in one direction (there is a single graviton escaping) the missing energy
distribution is harder than it would be in the case with two invisible particles
in the final state (which is the case for KK pair production).

\begin{figure}[t!] 
\centerline{
   \includegraphics[height=3.in]{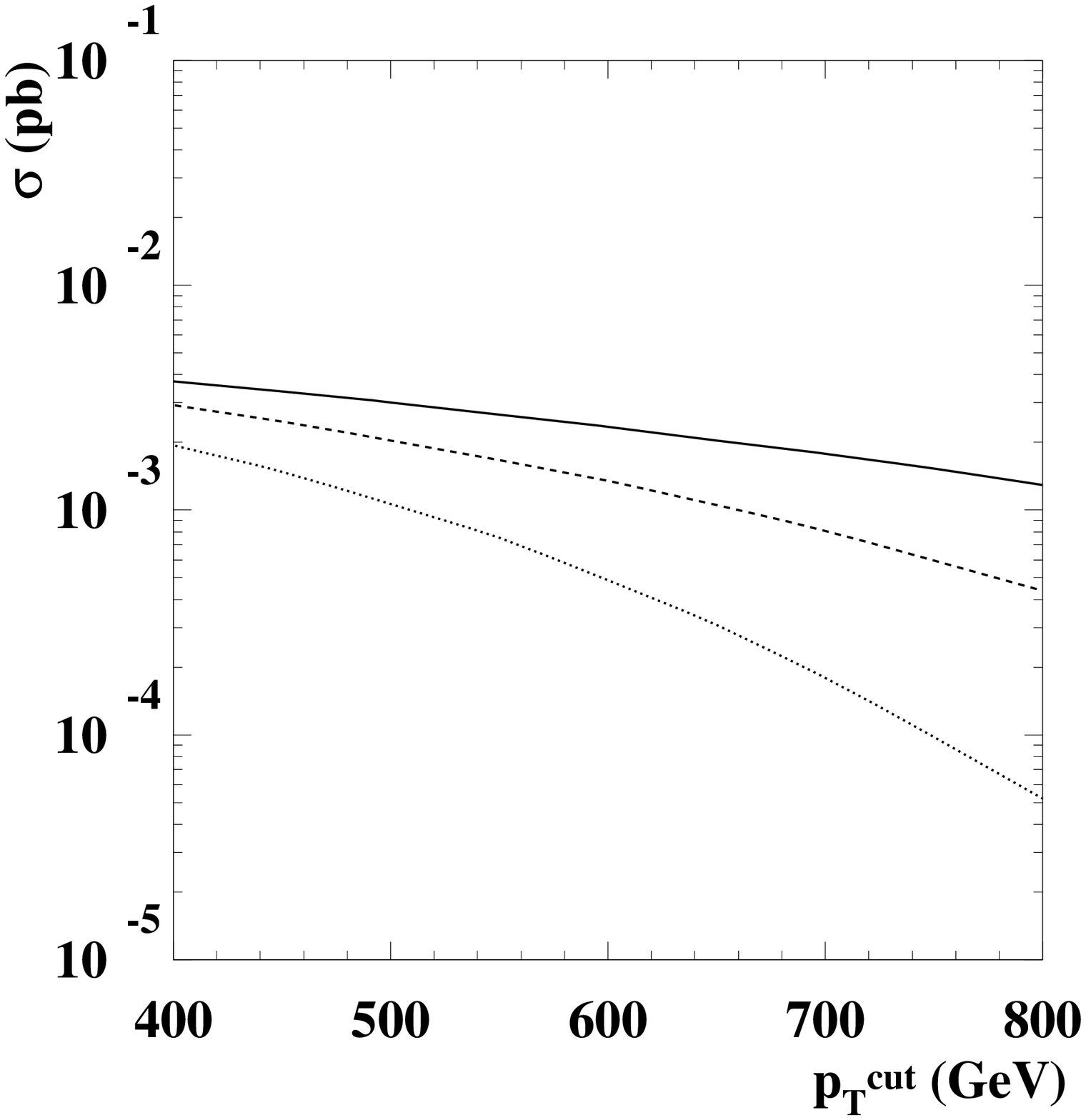}
   \includegraphics[height=3.in]{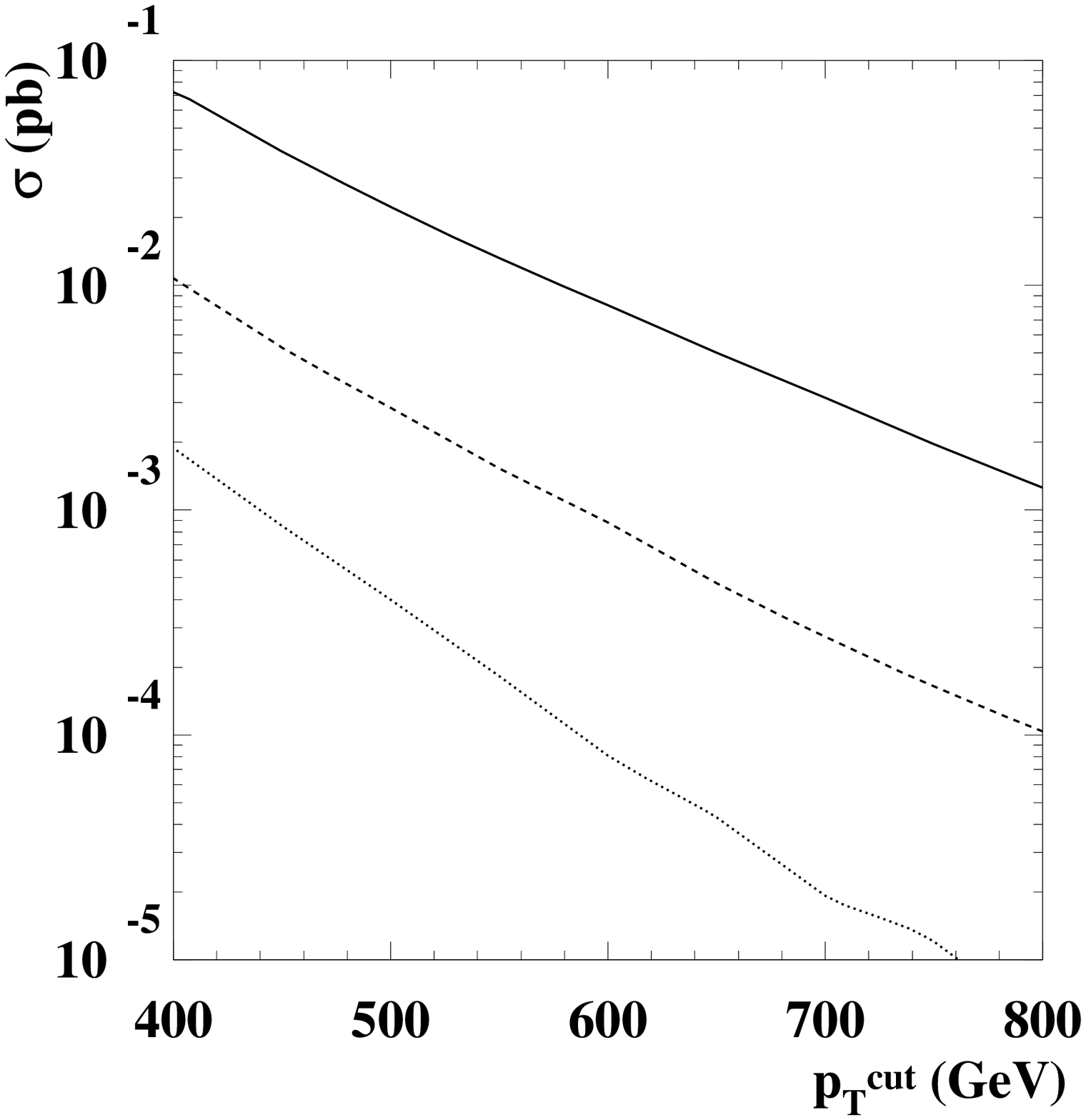}
   }
\caption{Signal(left) and backround(right) dependence on the parameter
$p_T^{\hbox{cut}}$. The three lines correspond to n = 1 (straight), 2 (dashed)
and 3 (dotted line). Cuts on rapidity and jet separation are also implemented.}
\label{bckg_cuts}
\end{figure}

We can assume then that by 
requiring a large transverse momentum and an even larger
missing transverse energy one will be able to get a good signal/background
ratio. We will use the following type of cuts: 
$p_{1T}, p_{2T} > p_T^{\hbox{cut}}, \not{E_T} >n p_T^{\hbox{cut}}$, with
n = 1, 2, 3. We also include a cut on rapidity $|y| < 4$ in our analysis, 
and we require that 
the two observable jets be separated by a cone with $R = \sqrt{\Delta \phi^2
+ \Delta \eta^2} > 0.4$. In Fig. \ref{bckg_cuts}(left) 
we plot the signal as a function of $p_T^{\hbox{cut}}$,
for $M = 3\ \hbox{TeV}, M_D = 10\ \hbox{TeV}$ and $N = 6$.

With such large cuts on the transverse momentum and missing energy, the dominant
backgrounds come from Z + 2 jets production and possibly QCD processes. Previous
estimates of this background \cite{tata_lhc} seem to indicate that for n = 1
and  $p_T^{\hbox{cut}} > 400$ GeV the 
magnitudes of these two contributions are about the same. However, one might 
suspect that as we increase the magnitude of the cut on the missing energy, the 
backgrounds due to mismeasurement in QCD processes will decrease, and
the dominant one will become the Z + 2 jets. This reasoning is supported by the
analysis in \cite{Gaines}, which at large missing energy 
shows a faster drop in the $\not{E_T}$ distribution
for QCD processes than for the Z + 2 jets.
  We therefore assume the the Z+ 2 jets 
process to be our entire background, and we evaluate it at parton level 
with the help of the MadEvent  generator \cite{madevent}.
The dependence of the background on the parameter $p_T^{\hbox{cut}}$ is plotted in Fig. \ref{bckg_cuts}(right), with the same 
kinematical cuts  as used for the signal.

\begin{figure}[t!] 
\centerline{
   \includegraphics[height=3.in]{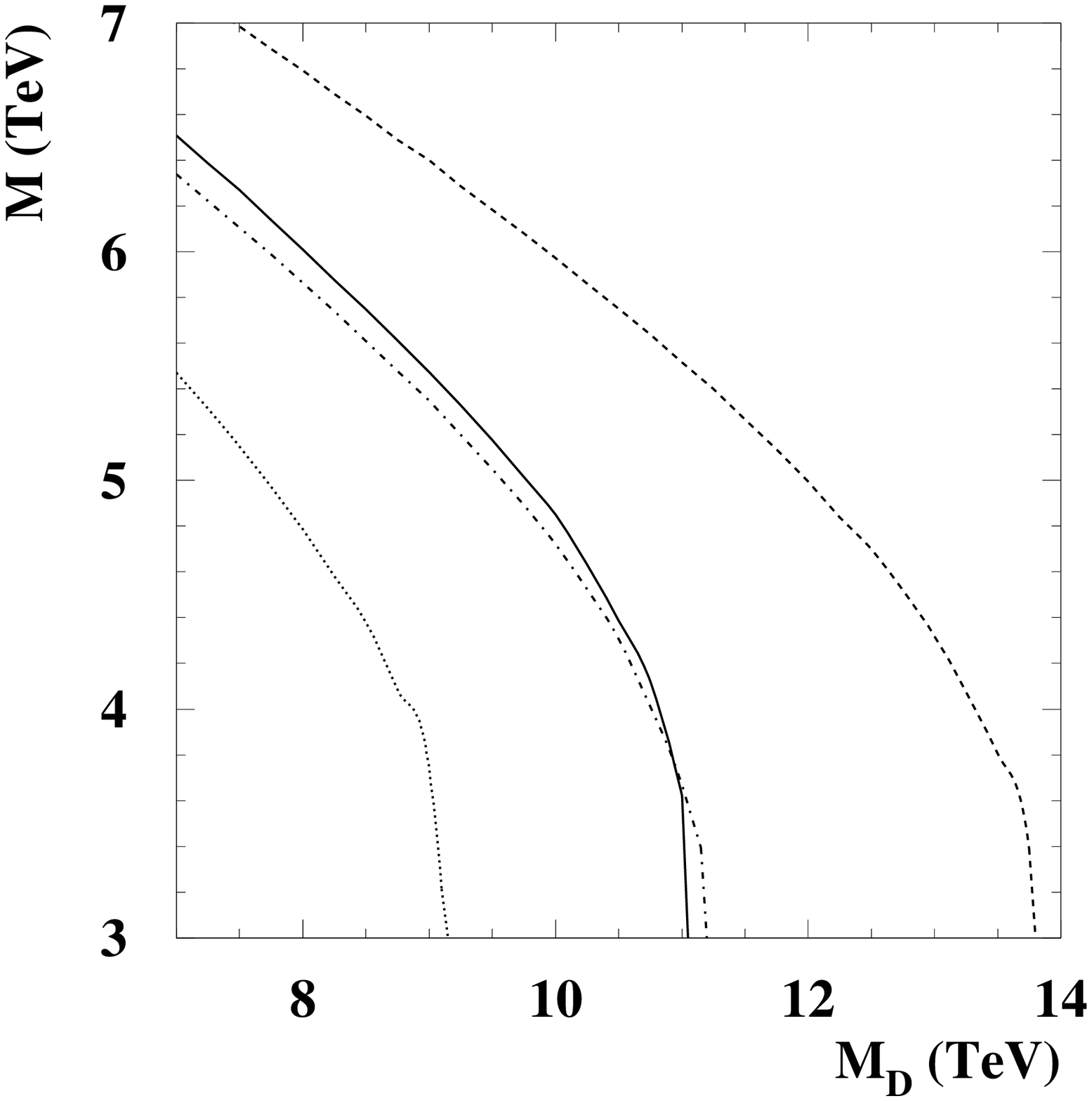}
   \includegraphics[height=3.in]{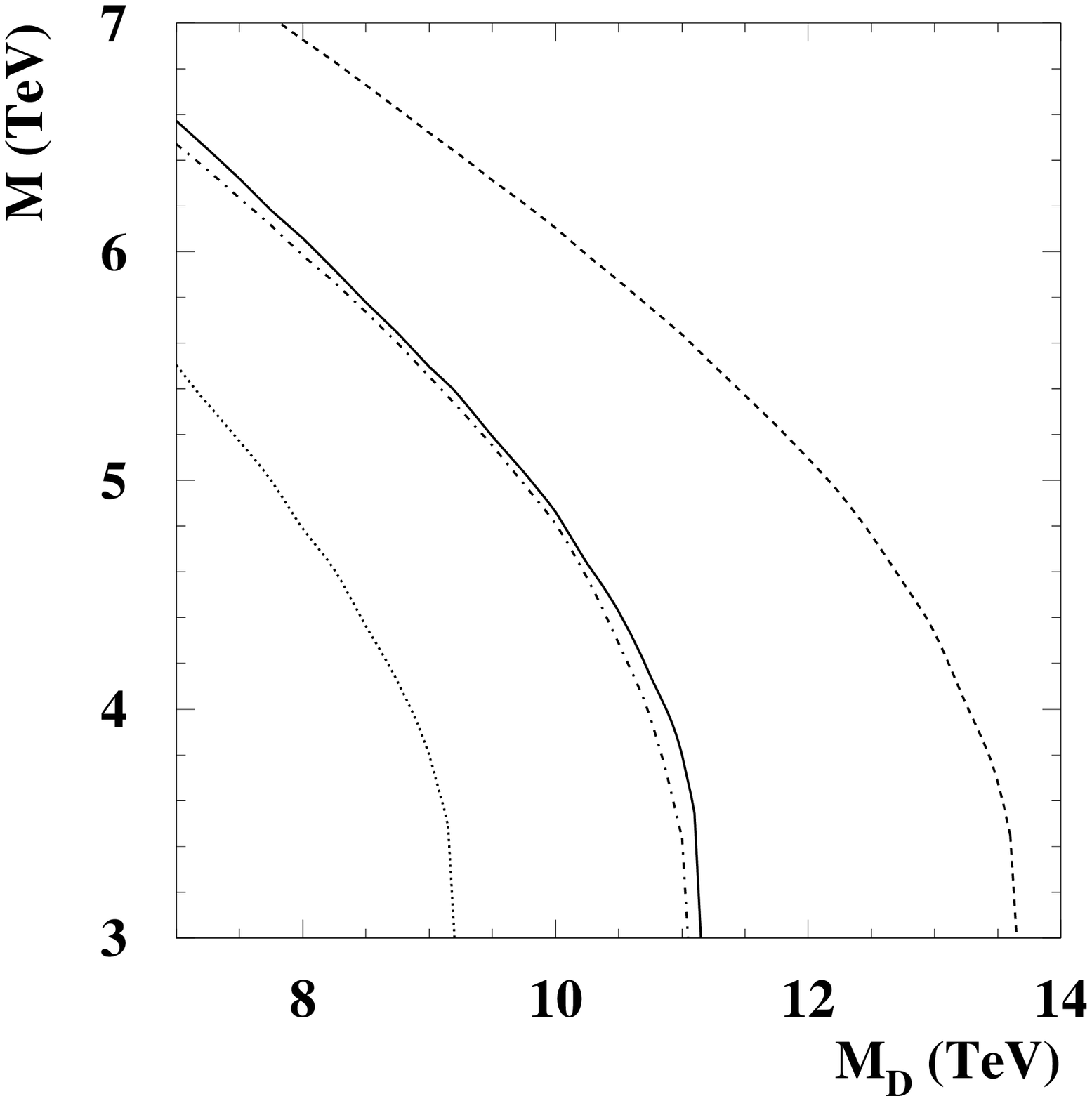}
   }
\caption{LHC reach for 20 and 100 signal events (straight and dotted lines 
corresponds to $N = 6$, and dashed and dotdashed line correspond to $N = 2$,
respectively.) Cuts a) are used in the left panel, and b) in the right panel.}
\label{lhc_reach}
\end{figure}

From this figure one can see that taking either one of the
following choice of cuts: 
a) $p_T^{\hbox{cut}} = 600$ GeV and $n = 3$, or b)
$p_T^{\hbox{cut}} = 800$ GeV and $n = 2$, the background will be $\sim 10$
events for an integrated luminosity of 100 fb$^{-1}$ at LHC. 
In Fig. \ref{lhc_reach} we plot the 
contour lines in the $M - M_D$ plane corresponding to the signal being 20
events (straigh line for $N = 6$ and dashed line for $N = 2$) and 100 events
(dotted line for $N = 6$ and dotdashed line for $N = 2$) at the same 
luminosity. The left plot in the 
figure was obtained  using the a) set of cuts, while the right plot was obtained
using the b) set of cuts. As noted before, the reach in $M$ can be 
large ( $\sim 6 $ TeV) for low values of $M_D$, 
but it drops quite fast as $M_D$ decreases.
We see from this plot that the most favorable case corresponds to $N = 2$, due 
to the fact that the production cross-section is larger, and also that
the transverse momenta of the jets are stronger in this case.

\begin{figure}[t!] 
\centerline{
   \includegraphics[height=3.in]{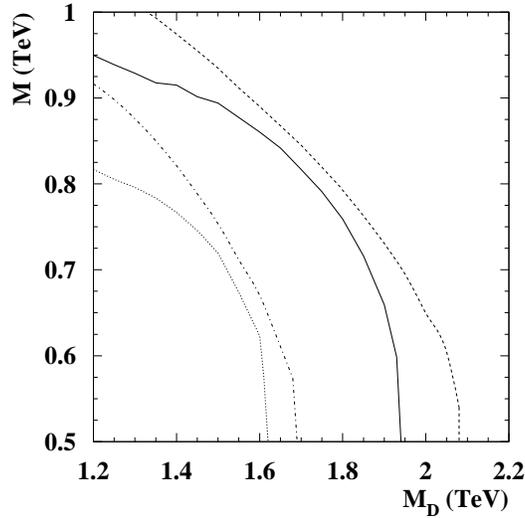}
   }
\caption{Tevatron Run II reach for 10 and 50 signal events 
(straight and dotted lines 
corresponds to $N = 6$, and dashed and dotdashed line correspond to $N = 2$,
respectively.) Here we use cuts with $p_T^{\hbox{cut}} = 150$ GeV and n = 2.}
\label{tev_reach}
\end{figure}

One can do the same type of analysis for the Tevatron RunII. We find that a
reasonable choice for the cuts on the jet transverse momenta and missing
energy is $ p_{1T}, p_{2T} > 150$ GeV, $\not{E_T} > 300$ GeV. With these
cuts the Z + 2 jets background (which is the dominant one here, 
too \cite{tata_tev}) is around one event, for an integrated luminosity of 
2 fb$^{-1}$. In Fig. \ref{tev_reach} we give the reach of the Tevatron
Run II for 10 and 50 signal events, for the number of extra dimensions in which
gravity propagates $N = 2$ and 6.  

\section{Conclusions}

We discuss the gravity mediated production of single KK excitations 
of gluons and quarks
at the Tevatron Run II and LHC. In standard Universal Extra Dimensions models, 
KK number conservation requires these excitations to be produced in pairs, 
thereby  requiring one  to pay a large price in phase space.
However, in our scenario, where
matter fields propagate on a fat brane, the gravitational interaction does not
obey KK number-conservation rules, allowing the production of a single KK
excitation and increasing the possible reach in mass. 
 
Thus, while for pair-produced KK excitations the LHC reach was found to be
around 3 TeV \cite{mmn}, for gravity mediated single KK production the
LHC reach can be as large as 7 TeV. However, the production
cross-section is strongly dependent on the fundamental gravity scale $M_D$.
We find out that in order for the production cross-section to be observable, 
the gravity scale $M_D$ should not be much larger than the  
scale $M$ which determines the mass of the matter KK excitations.

From an experimental viewpoint,
what results  is a picture in which one should look either for pair production
of low mass KK excitations ($M = 1/R \lesssim 3$ TeV), or production of a single KK excitation for
a higher value of the compactification scale $1/R$. The experimental signal
will be two jets plus missing energy in both cases, but there will be 
subtle differences in the distribution of the physical observables:
for example the missing energy distribution is likely to be harder in the
later case. Adding the possibility of additional decay channels
for the matter KK excitations (see, for example \cite{mmn2}) will further
enhance the differences between the signal in the two cases, 
and enrich the phenomenology of KK matter production.

\section*{Acknowledgments}
 We are grateful to U. Baur and X. Tata
for useful discussions. 
S. Nandi wishes to thank the Fermilab Theory Group for warm
hospitality and support during the completion of this work.
This work was supported in
part by the U.S. Department of Energy Grant Numbers
 DE-FG02-04ER41306, DE-FG02ER46140, and DE-FG02-85ER40231.


\end{document}